\documentclass{aa}

\def\Msun{\mbox{${\rm M}_{\odot}$}}

\def\Rsun{\mbox{${\rm R}_{\odot}$}}

\def\kms{\mbox{km\,s$^{-1}$}}

\usepackage{graphicx}
\usepackage{txfonts}
\usepackage{hyperref}
\usepackage{xcolor}

\newcommand{\spec}[3]{\mbox{#1\,{\sc #2}\,$\lambda$#3}\xspace}

\begin{document}

   \title{Spatially resolved centrifugal magnetosphere caught in motion around the secondary component of $\rho$ Oph A}
   \titlerunning{Resolved magnetosphere in motion around $\rho$ Oph Ab}

   \author{R.\ Klement
          \inst{1}
          \and
          M. E. Shultz
          \inst{1}
          \and
          Th.\ Rivinius
          \inst{1}
          }

   \institute{European Organisation for Astronomical Research in the Southern Hemisphere (ESO), Casilla 19001, Santiago 19, Chile\\
              \email{robertklement@gmail.com}
             }


 
  \abstract
   {}
   {The recently discovered spectroscopic binary $\rho$\,Oph\,A is a rare magnetic hot binary system composed of two early B-type stars, comprising a non-magnetic primary (Aa) and a slightly less massive magnetic secondary (Ab). Using near-IR interferometry, we aim to resolve the system's astrometric orbit. The high-spectral resolution VLTI/GRAVITY data also enables obtaining further information about the two stellar components, and about the centrifugal magnetosphere orbiting the magnetic star.}
   {We obtained a time-series of $K$-band interferometric data from VLTI/GRAVITY, and analyzed them with the geometrical model-fitting tool PMOIRED. The continuum was used to derive the relative astrometry and relative fluxes of the two components, leading to the astrometric orbital solution. Spectro-interferometry covering the Br$\gamma$ line was then used to obtain high-angular-resolution information about the dominant magnetospheric cloud in corotation with the magnetic Ab component.}
   {The combination of the astrometric orbital solution with the published radial velocities for both components led to a 3-dimensional orbital solution, dynamical masses, and a dynamical parallax, revealing a good agreement with previous estimates and confirming the previously inferred alignment of the orbital axis and the rotation axis of the Ab component, albeit at a much higher precision. Detailed analysis of the Br$\gamma$ spectrointerferometry then revealed the changing position of the magnetospheric cloud relative to the Ab component. Comparison to the location of the cloud predicted from $\rho$\,Oph\,Ab's oblique rotator model and H$\alpha$ emission properties demonstrated excellent agreement. We are additionally able to demonstrate that the magnetic star exhibits prograde rotation. This represents the first time that the motion of a centrifugal magnetosphere has been  detected in angularly resolved observations.}
   {}


   \maketitle
%

\section{Introduction}

Magnetic OBA stars comprise approximately 10\% of the population of the upper main sequence \citep{2017MNRAS.465.2432G,2019MNRAS.483.2300S}. Their surface magnetic fields are, with rare exceptions, geometrically simple \citep[approximately dipolar, e.g.][]{2018MNRAS.475.5144S,2019A&A...621A..47K} and strong \citep[with surface dipole strengths on the order of several kG, e.g.][]{2019MNRAS.483.3127S,2019MNRAS.490..274S}. They are also, without exception, stable: while there is clear and abundant evidence that the surface magnetic field strength and perhaps the total magnetic flux decline over evolutionary timescales \citep{land2007,land2008,2016A&A...592A..84F,2019MNRAS.490..274S}, over human timescales their magnetic fields show no evidence of reorganization \citep{2018MNRAS.475.5144S}. 

Since energy is transported by radiation rather than convection in the outer envelopes of OBA stars, contemporaneous dynamos are not expected. While magnetohydrodynamic (MHD) simulations predict that rotational-convective dynamos should produce extremely strong fields (hundreds of kG) in the convective cores of these stars \citep{2016ApJ...829...92A}, these should not be able to rise to the stellar surface on evolutionary timescales \citep{2024A&A...691A.326H}. It is therefore unlikely that dynamos sustained by core convection generate measurable hot star magnetic fields, and indeed the magnetic fields of hot stars show no correlation with surface rotation \citep{2019MNRAS.483.3127S,2019MNRAS.490..274S}, as would be expected for dynamo fields, and as is observed for cool stars \citep[e.g.][]{2016MNRAS.457..580F}. All of these factors lead to the characterization of hot star magnetic fields as `fossil' fields, i.e.\ remnants of some previous event in the star's life-cycle, and indeed, while radiative envelopes cannot generate magnetic fields, MHD simulations indicate that once established, they can be preserved almost indefinitely \citep[e.g.][]{2004Natur.431..819B,2010ApJ...724L..34D}. 

An important consequence of the stability of fossil magnetic fields is that the spectropolarimetric, spectroscopic, and photometric variability of magnetic early-type stars is dominated by rotation. Thus, in contrast to other hot stars, rotational periods are easily determined, frequently to a precision of around $10^{-6}$~d. The most common geometric configuration is a dipole tilted with respect to the rotational axis. This tilt of the magnetic dipole leads to spectropolarimetric variability due to the changing projection of the surface magnetic field as the star rotates. In those stars with detectable magnetospheres, this also leads to spectroscopic and photometric variability modulated on the rotational cycle, since the magnetosphere corotates with the star.

One possible avenue for generating fossil fields is via dynamos powered by binary mergers. This scenario has been demonstrated in MHD simulations \citep{2019Natur.574..211S}, and accounts for various aspects of the hot star population: binarity is extremely common on the upper main sequence; binarity is extremely rare among magnetic stars \cite{2015IAUS..307..330A}; the expected merger fraction is similar to the fraction of hot stars with fossil fields \citep{2012Sci...337..444S,2013ApJ...764..166D,2014ApJ...782....7D}. 

One phenomenon the merger scenario has some difficulty accounting for are the rare magnetic binary systems \citep[e.g.][]{2017A&A...601A.129L,2018MNRAS.478.1749K,2018MNRAS.475..839S,2019MNRAS.490.4154S}, since these would require initially triple or quadruple architectures. While mergers in such systems are by no means unexpected \citep{2024ARA&A..62...21M}, obtaining a tightly bound binary as the product of e.g.\ the merger of a hardened inner binary within a hierarchical triple is hard to explain \citep{2025A&A...693A..14B}. Magnetic binaries are therefore objects of particular interest in the study of hot star magnetism, as in addition to the usual advantages of binary stars (e.g. the precise measurement of fundamental parameters), they may ultimately offer some insight into the origin of fossil magnetic fields, and the evolutionary pathways open to the stellar population's brightest members. 

The nearby magnetic B2 star $\rho$\,Oph\,A (HD\,147933) was recently discovered by \citet{shultz_et_al_2025} to be a previously undetected spectroscopic binary, with two rapidly rotating B-type stars bound in a slightly eccentric 88-day orbit. The system consists of a non-magnetic primary, $\rho$\,Oph\,Aa, and a slightly less massive magnetic secondary, $\rho$\,Oph\,Ab. The system's architecture is strikingly similar to that of W\,601, a somewhat younger, slightly longer-period binary also consisting of two rapidly rotating B-type stars, with a magnetic secondary orbiting a non-magnetic primary \citep{2021MNRAS.504.3203S}.

The rotational variability of spectropolarimetric, spectroscopic, and photometric time series data all converge on a rotational period of around 0.74~d \citep{2020MNRAS.493.4657L}, making $\rho$\,Oph\,Ab one of the most rapidly rotating magnetic B-type stars known. The precisely measured rotational period enables the inclination angle of the rotational axis to be determined, with the result that this is apparently in alignment with the orbital axis (as constrained via comparison of the mass function to masses from evolutionary models). 

Due to its strong magnetic field and extreme rotation, $\rho$\,Oph\,Ab is surrounded by a `centrifugal magnetosphere' (CM), that is, a region in which the stellar wind is magnetically confined and forced into corotation with the star by the Lorentz force, with the resulting rotational support preventing gravitational infall of the confined plasma \citep{town2005c,petit2013}. CM plasma builds up to very high densities, before being expelled from the star in centrifugal breakout (CBO) magnetic reconnection events \citep{ud2008,2020MNRAS.499.5366O}. This results in several diagnostics that probe different parts of the magnetosphere. Electrons accelerated to semi-relativistic energies by CBO lead to gyrosynchrotron radio emission \citep{1987ApJ...322..902D,2021MNRAS.507.1979L,2022MNRAS.513.1429S,2022MNRAS.513.1449O} and beamed electron-cyclotron maser emission \citep{2000A&A...362..281T,2011ApJ...739L..10T,2021ApJ...921....9D,2022MNRAS.517.5756D}. The cool, pre-breakout plasma is detectable via photometric eclipses  \citep[when the magnetosphere passes in front of the star. e.g.][]{lb1978,town2008,2023MNRAS.523.6371B} as well as via H$\alpha$ emission \citep[when the magnetosphere is projected to the side of the star e.g.][]{grun2012,rivi2013,2015MNRAS.451.2015O,2015MNRAS.451.1928S,2016MNRAS.460.1811S,2020MNRAS.499.5379S,2021MNRAS.504.3203S}. All of these phenomena are present in the case of $\rho$\,Oph\,Ab. The H$\alpha$ emission of $\rho$\,Oph\,Ab is notably lopsided, with a single large cloud rather than the two clouds that are more commonly seen in single magnetic stars. The only other stars with such asymmetrical emission profiles are also close binaries \citep{2018MNRAS.475..839S,2020MNRAS.499.5379S}, suggesting that dynamical effects may play a role in shaping the magnetospheric plasma distribution. 

Since $\rho$\,Oph\,A is both bright ($V=5.05$) and very close (about 140 pc from the Sun), the predicted angular separation of several milliarcseconds (mas) puts it well within the angular resolution of the Very Large Telescope Interferometer (VLTI), thereby offering an excellent opportunity to precisely measure the orbital inclination, test the hypothesis that the rotational and orbital axes are aligned, and obtain dynamical measurements of the masses of the two stars. 

While the primary aim of the VLTI/GRAVITY $K$-band data presented in this work was to obtain astrometric constraints on the binary orbit, the high-spectral-resolution dataset also offers an opportunity to examine the magnetosphere in the Br$\gamma$ line, and to search for its spectro-interferometric signatures. To this date, the only interferometric detection of magnetically bound plasma was reported from a single observation of HR~5907 by VLTI/AMBER \citep{rivi2012b}.

The paper is structured as follows: in Sect.\ \ref{sec:observations} the VLTI/GRAVITY data are described. The orbital analysis is provided in Sect.\ \ref{sec:orbital_analysis}. The spectrointerferometric analysis of Br$\gamma$ is described in Sect.\ \ref{sect:brgamma}. The results are discussed in Sect. \ref{sect:discussion}, and the conclusions summarized in Sect.\ \ref{sect:conclusion}. Individual observations are provided in an Appendix (Section \ref{appendix}). 


\section{VLTI/GRAVITY Observations}\label{sec:observations}

\begin{figure*}[]
\begin{center}
\includegraphics[width=17cm]{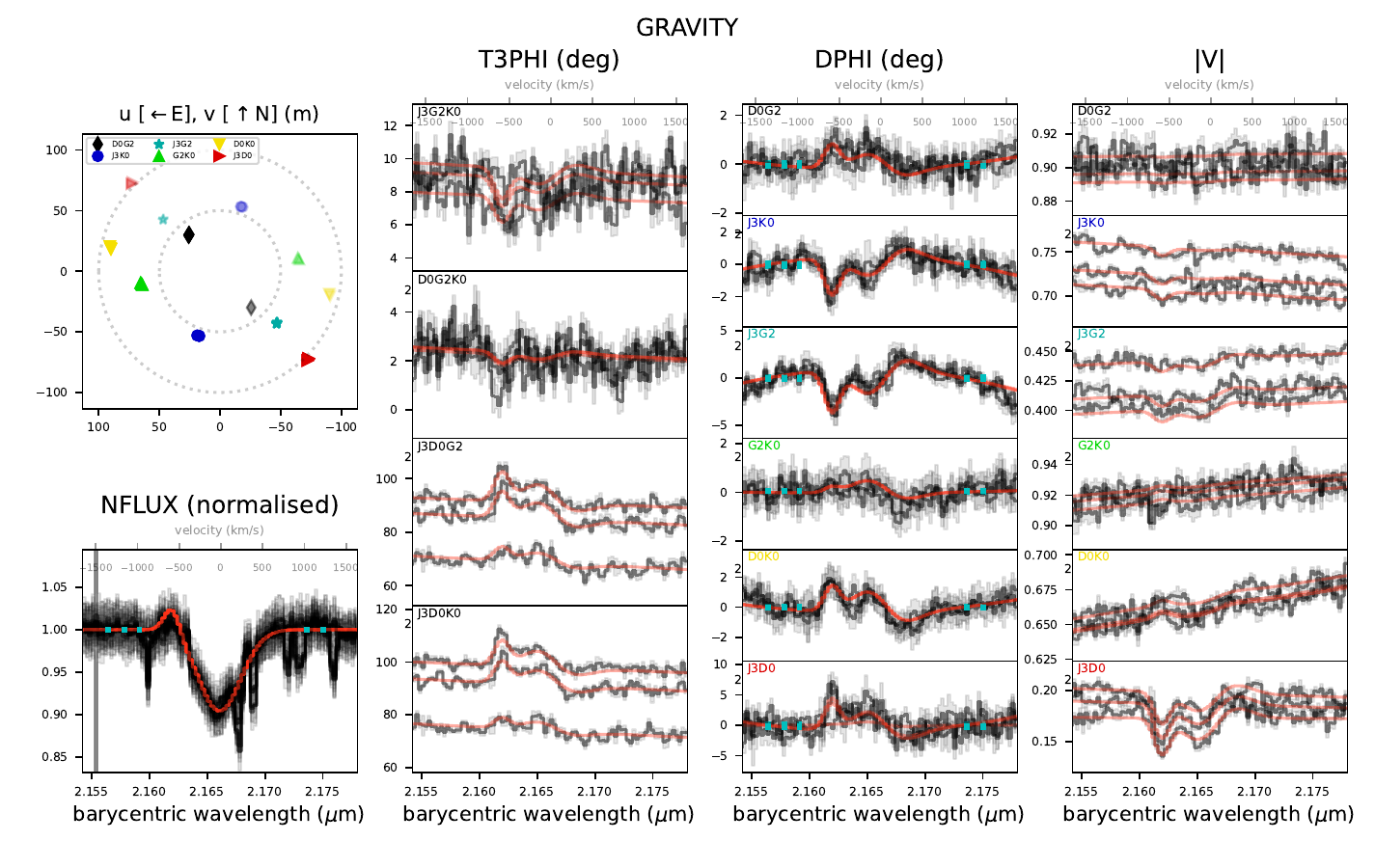}%
\end{center}
\caption[xx]{\label{fig:linefit}  The PMOIRED geometrical model fit to the Br$\gamma$ region of GRAVITY data taken on RJD 60139. The upper left panel shows the $(u,v)$ coverage, the lower left panel the normalized Br$\gamma$ line profile (NFLUX), and the remaining panels the interferometric observables closure phase (T3PHI), differential phase (DPHI), and the absolute visibility (|V|). The data (black with gray error bars) are overplotted with the model fit (red). The blue dots in the NFLUX and DPHI panels represent the continuum region used for normalization. The baselines and telescope triangles are given in the corner of each panel and their colors correspond to the $(u,v)$ coverage panel.}
\end{figure*}

We used the Very Large Telescope Interferometer \citep[VLTI,][]{2023ARA&A..61..237E} instrument GRAVITY \citep{2017A&A...602A..94G} operating in the near-IR $K$ band to collect a total of seven interferometric observations of the $\rho$~Oph~A binary system. VLTI was used with the Auxiliary Telescope (AT) Array in the `Large' configuration (as of 2023 Apr to 2023 Sep), reaching maximum baselines of $\sim130$\,m, which corresponds to the angular scale $\lambda / 2B_{\rm max} \sim 1.75$\,mas in the $K$ band. The four ATs have primary mirror sizes of 1.8\,m and are equipped with the dedicated adaptive optics system NAOMI \citep{2019A&A...629A..41W}. An example observation zoomed in on the Br$\gamma$ line is shown in Fig.\ \ref{fig:linefit}, with the different telescope baselines and triangles specified by the corresponding telescope locations on the VLTI platform in the upper left panel. For the layout of the telescope locations, see Fig.~10 in the VLTI manual\footnote{\url{https://www.eso.org/sci/facilities/paranal/telescopes/vlti/documents/VLT-MAN-ESO-15000-4552_v115.pdf}}.

The GRAVITY instrument was used in the single field on-axis mode, and in high spectral resolution ($R=4000$) covering the entire $K$ band ($1.98$ to $2.40$\,$\mu$m). Every interferometric `snapshot' ($\sim$1\,hr observing time including overheads) consisted of observation of the science target and the calibrator star HD\,145836 ($K = 5.1$, uniform disk diameter $=0.469\pm0.011$\,mas, spectral type K0\,III). For the reduction and calibration of the data, we used the dedicated European Southern Observatory (ESO) GRAVITY pipeline version 1.6.6 and the Eso\-Reflex software \citep{2013A&A...559A..96F}. The interferometric observables from GRAVITY covering the $K$ band include spectra (bottom left panel of Fig.\ \ref{fig:linefit}), calibrated closure phases (T3PHI, one for each telescope triangle per observation; second column of Fig.\ \ref{fig:linefit}), differential phases (DPHI, one for each telescope pair per observation; third column), and calibrated visibilities (|V|, one for each telescope pair per observation; fourth column). Each of the individual science and calibrator observations consisted of an object - sky - object - object sequence, resulting in three sets of interferometric observables of the science target for each snapshot. Variation in the observables (particularly in |V|) is noticeable in the three observations within each snapshot due to the Earth's rotation causing changes in the projected baselines. The seven snapshots are of very good quality with the exception of the one taken on RJD (Reduced Julian Date) 60082, which was marked as poorly executed by the ESO observing staff. While the RJD 60082 observation is of noticeably poor quality compared to the other snapshots (see Fig. \ref{fig:linefit_2023May16}), it could still be used in our analysis with certain limitations. The reduced and calibrated data will be made available in the Optical Interferometry Database\footnote{\url{http://oidb.jmmc.fr/index.html}} \citep{2014SPIE.9146E..0OH}.

For the modeling and visualization of the GRAVITY data we used the Python code Parametric Modeling of Optical InteRferomEtric Data\footnote{\url{https://github.com/amerand/PMOIRED}} \citep[PMOIRED,][]{2022SPIE12183E..1NM}. PMOIRED enables geometrical model fitting to GRAVITY data by least-squares minimization to obtain the best-fit parameters and includes various capabilities such as grid searching for the relative positions of components in multiple stellar systems, and fitting of spectro-interferometric signatures with spectral line profiles if the data have sufficient spectral resolution. Other features include a bootstrapping algorithm for the determination of realistic parameter uncertainties and a correction of the GRAVITY spectra for telluric features. The telluric correction enables assessment of the accuracy of the wavelength calibration, which is known to be at least $\sim0.02$\% \citep{2023A&A...672A.119G}. The telluric fitting of the $\rho$~Oph~A data suggests an accuracy of $\sim$0.03\% ($\sim$9\,\kms) or better, but we conservatively assumed an error of $0.05$\% ($\sim$15\,\kms) for the RVs derived from GRAVITY data by adding this error quadratically to the bootstrapped error from the model fit.

\begin{figure}[t]
\begin{center}
\includegraphics[width=\hsize]{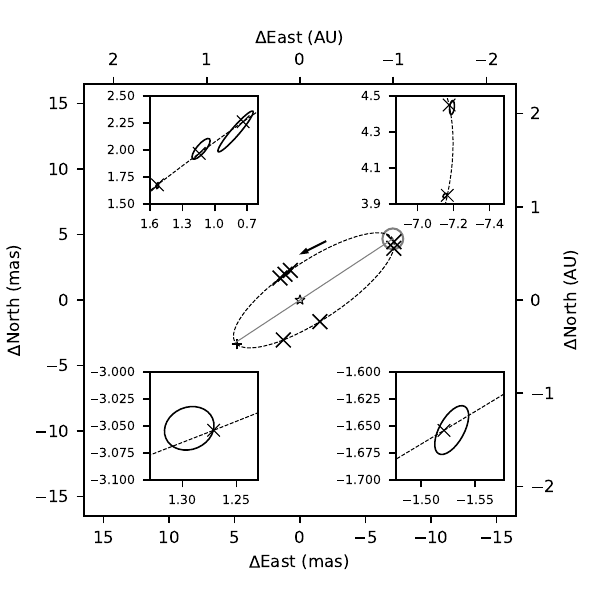}%
\end{center}
\caption[xx]{\label{fig:astrometric_orbit} The astrometric orbit (dashed line) with the primary non-magnetic Aa component at the center (star symbol), the line of nodes (gray line), the ascending node (grey empty circle), and the periastron position (plus sign). The calculated positions on the orbit for the observation epochs (x symbols) coincide with the astrometric measurements (error ellipses corresponding to 5-$\sigma$ uncertainties) which are too small to be seen in the main plot. The inset plots show the zoomed in areas around the measurements.
}
\end{figure}

\begin{figure}[t]
\begin{center}
\includegraphics[width=\hsize]{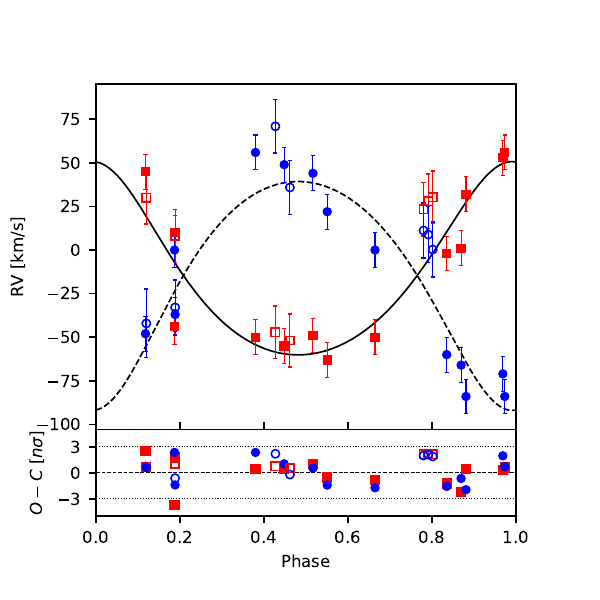}%
\end{center}
\caption[xx]{\label{fig:RV_curve} {\em Top}: the computed RV curves of the Aa and Ab components (solid and dashed curves, respectively) overplotted with RVs for the Aa and Ab components measured by Shultz et al. (filled red squares and blue circles, respectively), and RVs measured from the Br$\gamma$ signature in GRAVITY data (empty red squared and blue circles, respectively, see Sect.~\ref{sect:brgamma}).The latter RVs were not used for the orbital solution and are shown only for comparison. {\em Bottom}: Residuals in units of $\sigma$.
}
\end{figure}


\section{Orbital analysis}\label{sec:orbital_analysis}

To determine the relative astrometric positions from the seven GRAVITY snapshots, we used PMOIRED to fit the T3PHI and |V| observables with a simple geometrical model consisting of two uniform disks (UD) with fixed diameters of $0.3$ and $0.2$\,mas, representing the $\rho$~Oph~Aa and Ab components, respectively. The UD diameters of the two stars were estimated from the absolute radii and the distance determined by \citet{shultz_et_al_2025}, and are equivalent to unresolved point sources at the angular resolution of the data. The free parameters in the astrometric fitting are the astrometric offset $\rho$ of Ab relative to Aa, the astrometric position angle PA (measured from north to east), and the $K$ band flux ratio between the components $f=f_{\rm Ab}/f_{\rm Aa}$.

For the binary fitting, we selected the $K$-band wavelength interval between $2.10$ and $2.35$\,$\mu$m, and we rebinned the data from the original spectral resolution $R\sim4000$ to $R\sim100$. Using the grid search algorithm implemented in PMOIRED, the fit to all of the seven datasets successfully converged, resulting in $\rho$ ranging from $\sim$2.3 to $\sim$8.4\,mas, with reduced $\chi^2$ between 0.5 and 5.2 for the high quality epochs, and 9.4 for the low quality dataset from RJD 60082. An example of the rebinned data and the corresponding binary model fit is shown in Fig.~\ref{fig:binary_fit_2023Sep17}, and the resulting parameters, with the final values and uncertainties determined by bootstrapping, are given in Table~\ref{tab:astrometry}.

To obtain a 3-dimensional orbital solution, we combined the relative astrometry determined above with the radial velocities (RVs) measured by \citet{shultz_et_al_2025}. We used two different software packages for the orbital solution: the IDL code orbfit-lib\footnote{\url{https://www.chara.gsu.edu/analysis-software/orbfit-lib}} based on the Newton-Raphson method \citep{2016AJ....152..213S}, and the Python code spinOS\footnote{\url{https://github.com/matthiasfabry/spinOS}} with the implemented Markov Chain Monte Carlo (MCMC) method \citep{2021A&A...651A.119F,2021ascl.soft02001F}. The weights of the two datasets were adjusted so that they contribute approximately equally to the total residuals. The two resulting sets of orbital parameters, as well as the dynamical masses and the distance to the system, are fully consistent with each other (Table~\ref{tab:orbital_solution}). The spinOS MCMC method gives considerably smaller parameter uncertainties, except for the angular semimajor axis ($a"$), which for the spinOS solution has to be computed from the fitted total mass and distance. The results agree well with the spectroscopic orbital parameters given by \citet{shultz_et_al_2025}, which are provided for comparison in the final column of Table \ref{tab:orbital_solution}. The resulting astrometric orbit and RV curves computed from the orbfit-lib solution are plotted in Figs.~\ref{fig:astrometric_orbit} and \ref{fig:RV_curve}, respectively.

\begin{table*}[]
\caption[xx]{\label{tab:orbital_solution} Orbital solutions}
\begin{center}
\begin{tabular}{lccc}
\hline\hline
Parameter           & orbfit-lib & spinOS & S25\tablefootmark{a}\\
\hline
$P$ [d]              & $87.849\pm0.024$    & $87.831\pm0.010$  & $88.00 \pm 0.02$\\
$T_0$ [RJD]          & $60101.08\pm0.20$   & $60101.111\pm0.052$ & $53511.8 \pm 0.07$ \\
$e$                  & $0.17967\pm0.00070$ & $0.17931\pm0.00017$ & $0.13 \pm 0.05$\\
$a''$ [mas]          & $7.203\pm0.010$     & $7.20\pm0.27$ & --\\
$i_{\rm orb}$ [$^\circ$] & $71.311\pm0.072$ & $71.348\pm0.020$ & $69 \pm 11$\\
$\Omega$ [$^\circ$]  & $303.38\pm0.11$     & $303.385\pm0.034$ & $301 \pm 4$ \\ 
$\omega_{\rm Aa}$ [$^\circ$]  & $4.72\pm0.86$ & $4.80\pm0.24$ & $32\pm4$ \\
$K_{\rm Aa}$ [\kms]  & $55.2\pm6.3$        & $55.5\pm3.6$ & $57 \pm 5$ \\
$K_{\rm Ab}$ [\kms]  & $65.6\pm6.3$        & $65.7\pm3.6$ & $67 \pm 5$ \\
$\gamma$ [\kms]      & $-14.6\pm3.3$       & $-14.6\pm1.5$ & $-16 \pm 2$\\
\hline
$a$ [AU]                   & $1.014\pm0.074$ & $0.995\pm0.026$ & $1.06 \pm 0.08$    \\
$q=M_{\rm Ab}/M_{\rm Aa}$  & $0.84\pm0.12$   & $0.845\pm0.072$   &  $0.83 \pm 0.14$ \\
$M_{\rm total}$ [\Msun]    & $18.0\pm4.0$    & $17.0\pm1.3$   & $21 \pm 8$    \\
$M_{\rm Aa}$ [\Msun]       & $9.8\pm2.2$     & $9.21\pm0.79$  & $11 \pm 3$   \\
$M_{\rm Ab}$ [\Msun]       & $8.2\pm2.0$     & $7.79\pm0.70$  & $10 \pm 3$  \\
$d$ [pc]                   & $141\pm10$      & $138.2\pm3.6$  & $139 \pm 3$     \\
\hline
rms astrometry [$\mu$as] & $18$ & $19$ & -- \\
rms RV$_{\rm Aa}$ [\kms]  & $15.5$ & $15.9$ & $11.9$\\
rms RV$_{\rm Ab}$ [\kms] & $14.5$ & $15.5$ & $15.2$\\
\hline
\end{tabular}
\end{center}
\tablefoot{$P$ is the orbital period, $T_0$ is the epoch of periastron or the epoch of the superior conjunction, $e$ is the eccentricity, $a''$ is the angular semi-major axis of the orbit, $i_{\rm orb}$ is the orbital inclination, $\Omega$ is the longitude of the ascending node, $\omega$ is the longitude of the periastron, $K$ are the velocity semiamplitudes, $\gamma$ is the systemic velocity, $a$ is the orbital semi-major axis, $q=M_{\rm Aa} / M_{\rm Ab}$ is the mass ratio, $M_\mathrm{total}$ is the total mass, and $M_{\rm Ab}$ and $M_{\rm Aa}$ are the masses of the two components. `rms' stands for the root mean square residuals of specified datasets. \\
\tablefoottext{a} {Orbital elements and physical parameters based on the spectroscopic solution and evolutionary fitting from \citet{shultz_et_al_2025}.}
}
\end{table*}


\section{Spectro-interferometric analysis}\label{sect:brgamma}

\begin{figure}
\begin{center}
\includegraphics[width=\hsize]{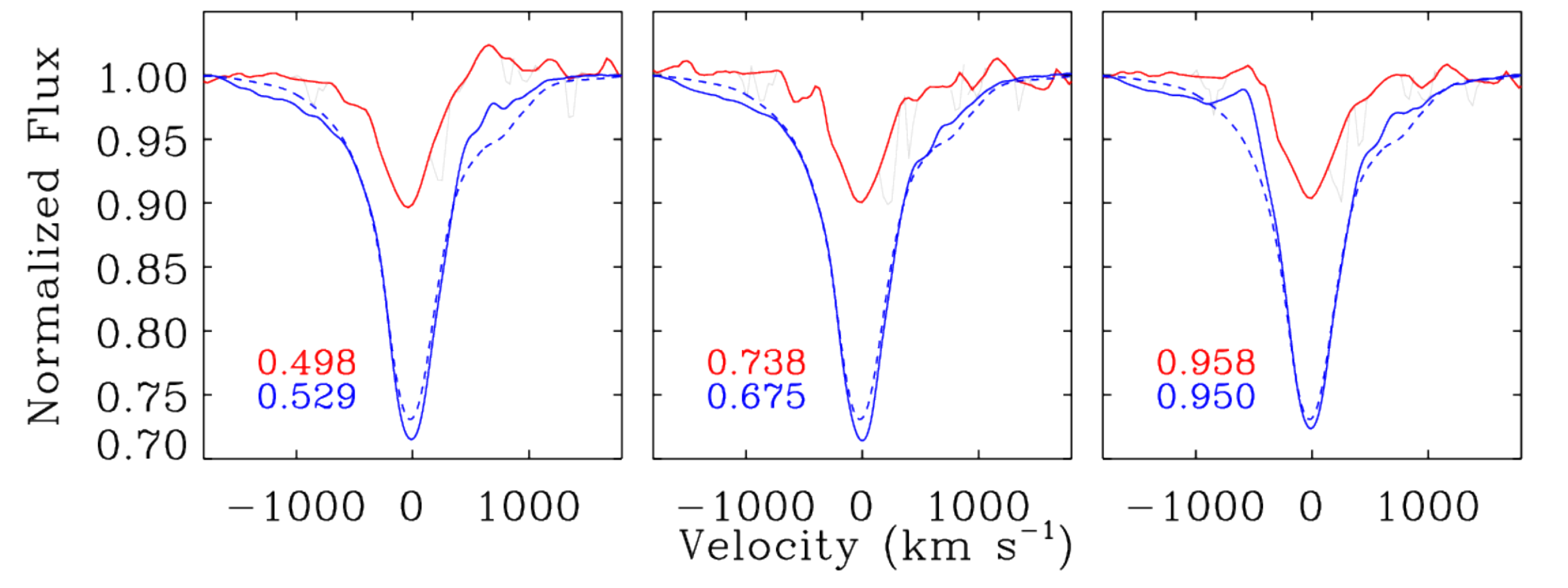}
\end{center}
\caption[]{Comparison between Br$\gamma$ (red) and H$\alpha$ (blue) at similar rotation phases (labels in lower left corners). The dashed blue line is a synthetic spectrum, helping to highlight the location of the H$\alpha$ emission. Note that emission is neither expected nor seen at near phase 0.7 (middle).}
\label{fig:rhoopha_halpha_brgamma}
\end{figure}

The relatively high spectral resolution of the GRAVITY data ($R=4000$) enables analysis of normalized spectral line profiles (NFLUX) in the $K$ band and their corresponding spectro-interferometric signatures in the T3PHI, DPHI, and |V| observables. In our data, spectro-interferometric signal is detected in two spectral lines: the hydrogen line Br$\gamma$ and \spec{He}{i}{21126}. For the geometrical model fitting of the spectro-interferometry described below, we used all of these four observables except for the low-quality epoch RJD 60082, for which we were forced to discard the T3PHI and |V| data (see below).

\subsection{Br$\gamma$}

The hydrogen line Br$\gamma$ is the most prominent line in the $K$ band spectra of $\rho$~Oph~A. Analogous to H$\alpha$ in the visible, Br$\gamma$ shows a blended, broad absorption profile from the two rapidly rotating binary components, and in all but one dataset also an emission bump from the magnetosphere. Fig.\ \ref{fig:rhoopha_halpha_brgamma} compares Br$\gamma$ to H$\alpha$ at similar rotational phases, demonstrating that emission is seen in Br$\gamma$ at the expected velocities. The emission bumps in these datasets are highly shifted in RV and located outside of the broad absorption profile, indicating that $\rho$~Oph~Ab was caught at rotational phases when the magnetosphere was projected next to the star. The seventh dataset taken on RJD 60171 (phase 0.738 in Fig.\ \ref{fig:rhoopha_halpha_brgamma}) shows no emission bump; this is expected, as H$\alpha$ does not display emission at this phase, which occurs when the magnetospheric cloud is eclipsed behind the star \citep[see][]{shultz_et_al_2025}.

Based on the Rigidly Rotating Magnetosphere \citep[RRM;][]{town2005c} model, the inner edge of the magnetospheric cloud should be located above the Kepler rotation radius $R_{\rm K} = 1.93^{+0.08}_{-0.03}~R_*$ \citep{shultz_et_al_2025}, a prediction consistent with the properties of the star's H$\alpha$ emission. Given the velocity range of H$\alpha$ at maximum emission, the outermost extent of the cloud is around $4.4$\,{\Rsun} (see below). At a distance of about 140 pc, the expected maximum projected angular extent of the magnetosphere -- occurring at rotational phases 0.0 and 0.5 -- should then be $\sim0.25$\,mas, and it should be separated from the center of the star at the same phases by $\sim0.2$ to $\sim0.45$\,mas. Given the high quality and angular resolution of the GRAVITY data, such an offset should introduce a small but detectable signal in the Br$\gamma$ DPHI ($\sim1^{\circ}$ at the longest baselines), similar to what was detected by \citet{rivi2012b} for HR~5907 in AMBER data. The detectability of the magnetospheric cloud is further aided by the fact that it surrounds the secondary component in a low contrast binary, which amplifies the expected DPHI signal, and introduces small but detectable signals in the T3PHI and |V| observables as well.

Visual inspection of the interferometric observables in the datasets with the magnetospheric emission bump reveals rather complex features (especially in DPHI), and there is indeed a clear signal at a wavelength identical to the emission bump in the spectrum (see Fig.\ \ref{fig:linefit}). In addition, there is a broader signal corresponding to the relative astrometric offset of the two stellar components, whose absorption profiles should also be RV-shifted following the orbital solution.

For the detailed analysis of the Br$\gamma$ line profile and spectro-interferometry, we selected a 24\,nm-wide region centered on the Br$\gamma$ vacuum wavelength $\lambda_{\mathrm{Br}\gamma\mathrm{, vac}} = 2.1661178$\,$\mu$m, and we employed PMOIRED to fit a geometrical model to the unbinned data ($R\sim4000$). We used the same representation for the two stellar components as in the binary fitting procedure, and we kept the relative astrometry ($\rho$ and PA) and the flux ratio ($f$) fixed at the previously determined values (Sect.~\ref{sec:orbital_analysis}, Table~\ref{tab:astrometry}). Then, we added the following model components: two Gaussian absorption line profiles representing the lines of the two stellar components, and a Gaussian emission line profile representing the emission bump from the magnetosphere. Each of these components is defined by the line FWHM and peak amplitude, the RV shift of the line center, and the astrometric position (again fixed at the previously determined values for the two stars). The astrometric position of the magnetosphere was allowed to converge from a starting position corresponding to the magnetic Ab component. While the RVs of the stellar components could have been fixed at the values derived from the orbital solution (Sect.~\ref{sec:orbital_analysis}), the GRAVITY data proved good enough to fit them, enabling an additional check of the validity of our orbital solution.

\begin{figure}[t]
\begin{center}
\includegraphics[width=\hsize]{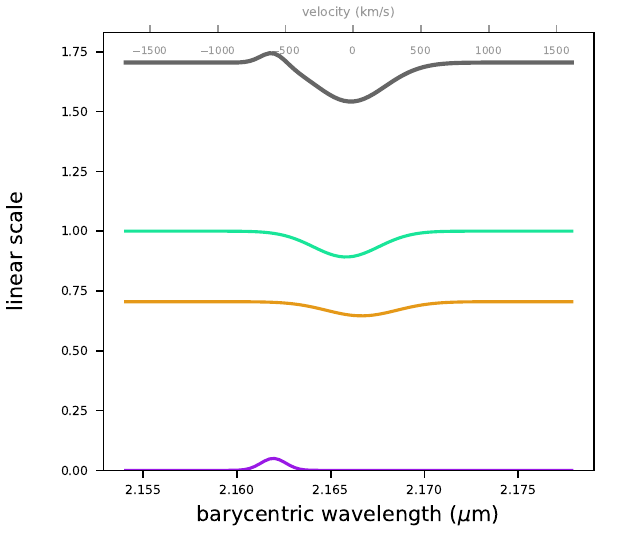}%
\end{center}
\caption[xx]{\label{fig:model_spectra} Model spectra obtained for the epoch of RJD 60139. Green spectrum is the Aa component with its continuum flux fixed to 1.0, orange spectrum is the Ab component with $f=f_{\rm Ab}/f_{\rm Aa}=0.71$, purple spectrum is the magnetospheric emission with no continuum contribution, and black spectrum is the sum of the three components. 
}
\end{figure}

\begin{figure}
\begin{center}
\includegraphics[width=0.5\textwidth]{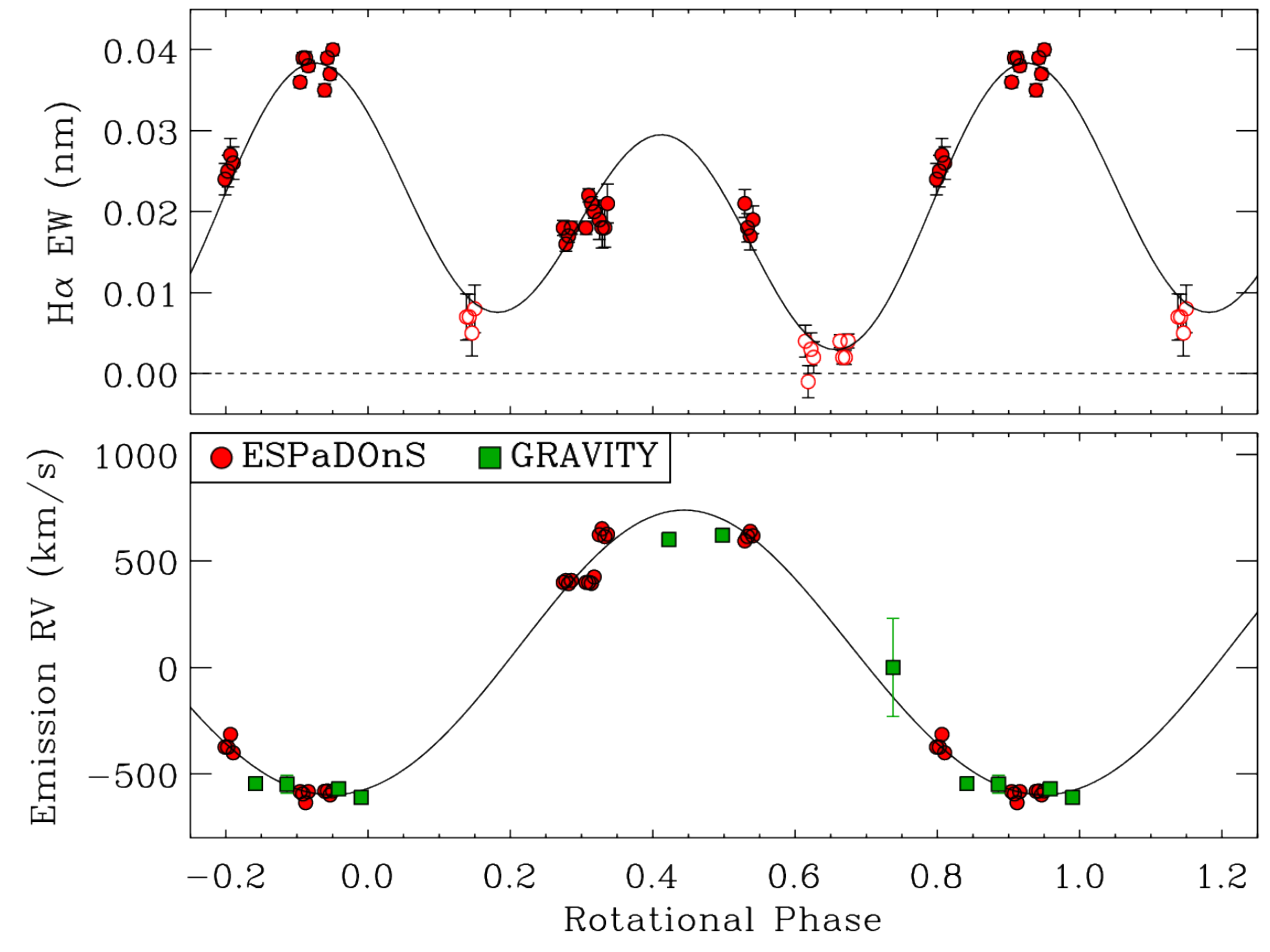}
\end{center}
\caption[]{{\em Top}: H$\alpha$ emission EW  folded with the rotational period, demonstrating the double-wave character of the variation. The approximate continuum level corresponds to the null line. Points with no emission (from which useful RVs cannot be measured) are indicated by open circles. {\em Bottom}: RVs measured from the emission peaks of H$\alpha$, and the RVs obtained from the Br$\gamma$ line fitting, both folded with the rotational period. In both panels the black curve shows a harmonic fit to the ESPaDOnS data.}
\label{fig:rhooph_halpha_peak_rv}
\end{figure}

\begin{figure*}
\begin{center}
\includegraphics[width=\textwidth]{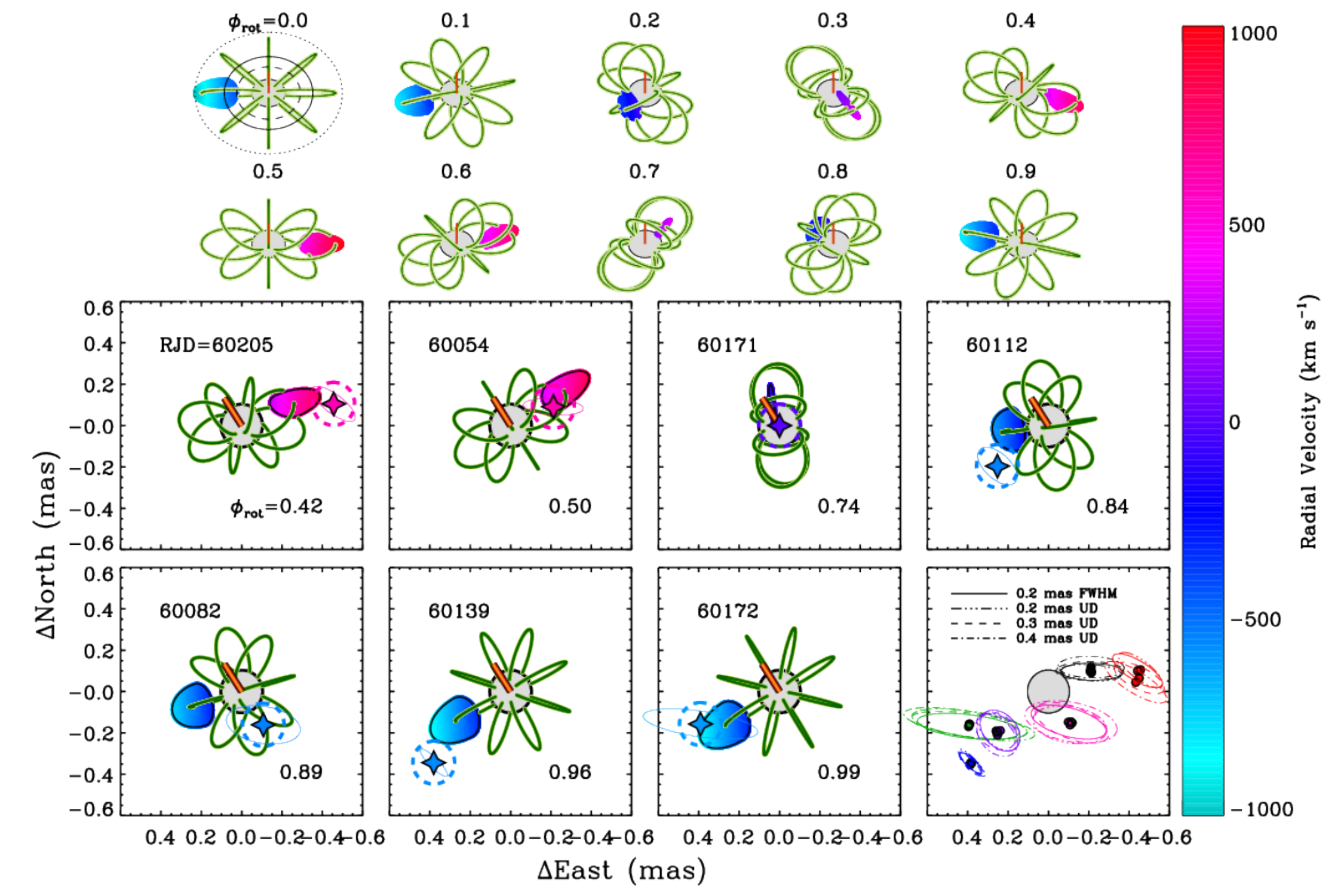}
\end{center}
\caption[]{{\em Top}: Model showing the projected position of the main magnetospheric cloud in rotational phase increments of 0.1. Cloud color indicates radial velocity, as indicated by the color bar. The rotational axis is shown by the orange line. In the top left, dashed, solid, and dotted lines respectively indicate the Kepler corotation radius, the radius of maximum emission, and the outer extent of emission. {\em Bottom}: Comparison between the expected location of the main magnetospheric cloud and the position recovered from the GRAVITY spectro-interferometry in Br$\gamma$. Astrometric locations are indicated by the star symbol; uncertainties in the position by error ellipses; the assumed angular size of the cloud used in the fitting by the dashed circle. {\em Bottom right}: comparison between cloud positions and error ellipses inferred from different models: a Gaussian model with an 0.2 mas FWHM, and uniform disk models with radii of 0.2, 0.3, and 0.4 mas. Individual observations are indicated by color. Adoption of different modelling assumptions has no significant influence on disk position or error ellipse properties.}
\label{fig:rhoopha_spectroinf_clouds}
\end{figure*}

The geometrical model described above proved to be a good representation of the Br$\gamma$ region for the five good quality datasets featuring the magnetospheric emission bump, namely the epochs RJD 60054, 60112, 60139, 60172, and 60205, with all of the free parameters (11 in total) successfully converging. An example of the fit is shown in Fig.~\ref{fig:linefit} (for the remaining epochs see Figs.~\ref{fig:linefit_2023Apr18} to \ref{fig:linefit_2023Sep17}), with the components of the corresponding model spectrum shown in Fig.~\ref{fig:model_spectra}. 

Regarding the lower quality dataset from RJD 60082, we were forced to discard the T3PHI and |V| observables due to the lack of any discernible spectro-interferometric signal in the noisy data. The DPHI part, on the other hand, shows a clear signal from the magnetosphere for three out of the six baselines, while for the other three the data are again too noisy. The DPHI and the line profile were successfully fitted after fixing the FWHM and peak amplitudes of the three line components at the averaged best-fit values determined from the other five epochs, with only the magnetosphere position and the three RVs left as free parameters. 

As for the RJD 60171 epoch, no discernible signal from the magnetosphere can be seen in Br$\gamma$, and we were able to fit only the two line components corresponding to the two stars. This is consistent with the expectation that at this rotational phase the magnetosphere is eclipsed by its host star. 

As with the binary positions determined in Sect.\ \ref{sec:orbital_analysis}, the final values of all the converged parameters and their uncertainties were obtained by the bootstrapping algorithm implemented in PMOIRED. The resulting RVs of the two stars and RVs and astrometric positions of the magnetospheric cloud are listed in Table~\ref{tab:gravity_rvs_cloud_astrometry}. 

The main goal of this analysis was to provide the astrometric positions and RVs of the magnetosphere, as well as RVs of the two stars for all seven epochs. With the exception of RJD 60171, the astrometric positions of the magnetosphere converged to a relative offset of a few tenths of mas from the Ab component, while the magnetosphere RVs were found to vary around $+600$\,{\kms} or $-600$\,{\kms}, matching the expectations (see below for a detailed comparison). Due to the lack of magnetospheric signal at RJD 60171, the magnetosphere position and RV could not be fitted, but we can safely assume that the magnetosphere position in this case coincides with the Ab component, while its RV is close to the systemic velocity; in this case we adopt an error bar of $\pm230$\,\kms, corresponding to the stellar $v \sin{i}$. The stellar RVs are shown in Fig.~\ref{fig:RV_curve} as empty symbols, and demonstrate a reasonable agreement with both the RVs measured by Shultz et al.\ and our orbital solution in spite of the broadness of the line profiles and the relatively low spectral resolution of GRAVITY (compared to high-resolution spectrographs). 

Fig.~\ref{fig:rhooph_halpha_peak_rv} compares the RVs measured for the Br$\gamma$ emission bump to the velocities of peak emission measured from H$\alpha$ using ESPaDOnS data. The top panel of Fig.\ \ref{fig:rhooph_halpha_peak_rv} shows the H$\alpha$ `emission equivalent widths' measured by Shultz et al., folded with the star's rotational period. These were obtained by measuring the equivalent width (EW) between about $\pm$1000~\kms~(the approximate span of the emission profile) of the observed H$\alpha$ line, and subtracting the EW of a synthetic binary spectrum (see Fig.\ \ref{fig:rhoopha_halpha_brgamma}). The emission EWs vary with a double-wave pattern, illustrated with a harmonic fit, with two local maxima corresponding to phases at which the magnetosphere is seen face-on, and local minima corresponding to phases at which the magnetosphere is either eclipsing or eclipsed by the star. As demonstrated by Shultz et al., these phases respectively correspond to the maxima and the nulls of the longitudinal magnetic field curve, i.e.\ phases at which the magnetic axis is closest to alignment with the line of sight, and phases at which the magnetic equator bisects the visible stellar disk. 

The bottom panel of Fig.\ \ref{fig:rhooph_halpha_peak_rv} shows the RVs of the H$\alpha$ and Br$\gamma$ emission peaks, also folded with the rotational period. A harmonic fit to the H$\alpha$ data follows the expected sinusoidal pattern, with the velocity of peak emission varying smoothly between $\pm$600~\kms. Open symbols, corresponding to emission EW minima (i.e. phases at which the cloud is either eclipsing or eclipsed), were not used in the fit, as the emission peak velocities at these phases are unreliable. Green squares indicate the velocities obtained from Br$\gamma$ spectro-interferometry. They follow the expected variation from H$\alpha$ very closely.

The positions of the magnetosphere relative to the Ab star are compared to the predictions of the rotational, magnetic, and magnetospheric model parameters determined by \citet{shultz_et_al_2025} in Fig.~\ref{fig:rhoopha_spectroinf_clouds}. The top panels of Fig.\ \ref{fig:rhoopha_spectroinf_clouds} show a schematic model of the projected location of the dominant magnetospheric cloud at various rotational phases. The inner edge of the cloud is placed at the Kepler corotation radius $R_{\rm K} = 1.9~R_*$ (dashed line in the upper left panel). The radius of maximum emission $r_{\rm max}$, shown by the solid line, was inferred from the RV of the H$\alpha$ emission line peak, $v_{\rm max} = -600$~\kms: since the cloud is in strict corotation with the star, RV maps linearly to projected distance from the center of the star such that $r_{\rm max} = v_{\rm max} / v\sin{i} = 2.6~R_*$. The outermost extent of the cloud $r_{\rm out} \sim 4.4~R_*$, shown by the dotted line, was similarly inferred from the RV at which the emission line disappears. The RV of the cloud is mapped to color, as indicated by the color bar.

The star's rotation axis is inclined at an angle $i_{\rm rot} = 71^\circ$ from the line of sight, which is taken from the orbital fit (Sect.~\ref{sec:orbital_analysis}) and matches the rotational inclination determined by Shultz et al. ($73 \pm 5^\circ$). The magnetic field, assumed to be a dipole, is then tilted from the rotational axis at an angle $\beta = 62^\circ$. The angle $\beta$ was determined as one of the oblique rotator model parameters inferred from magnetic modeling of the longitudinal magnetic field curve; the oblique rotator model parameters were verified via direct comparison to the circular polarization profiles \citep{shultz_et_al_2025}. As can be seen from the model, the cloud begins at phase 0.0 to the left of the star, at maximum negative RV and maximum projected area (corresponding to maximum emission, see Fig.\ \ref{fig:rhooph_halpha_peak_rv}). Since the cloud is in the magnetic equatorial plane, this phase also corresponds to a nearly pole-on view of the magnetic field. Between phases 0.2 and 0.3 the cloud eclipses the star, corresponding to the magnetic equator crossing the line of sight. The cloud then moves to the right of the star and positive RV, as the negative magnetic pole moves into the line of sight; passes behind the star at phase 0.7, corresponding to the second magnetic null; and then reappears with negative RV to the left of the star. 

The bottom two rows of Fig.\ \ref{fig:rhoopha_spectroinf_clouds} show the model at the rotational phases of the GRAVITY observations, comparing the expected position and RV of the cloud to the positions and velocities determined from spectro-interferometric modeling described above. Here the rotational axis has been rotated in the plane of the sky to be perpendicular to the longitude of the ascending node $\Omega = 303^\circ$, as should be the case if the spin and orbital axes are aligned. In general there is good agreement between observations and models. For the observations on RJDs 60205, 60054, and 60172, astrometric and model positions overlap within the 1$\sigma$ error ellipses. The error ellipse of the observation on RJD 60112 is just in contact with the expected outer extent of the cloud. The error ellipse of the observation of RJD 60139 is well outside of the expected cloud position, however the cloud size assumed in the astrometric modeling (dashed circle) indicates overlap with the cloud position, and the measured position angle matches the position angle of the cloud. 

Two observations deserve closer examination. First is the observation on RJD 60171, for which no cloud position could be determined, and no Br$\gamma$ emission could be detected. The astrometric position of the cloud shown in Fig.\ \ref{fig:rhoopha_spectroinf_clouds}, at the exact center of the star, is purely nominal, and there is no error ellipse. As can be seen from the model, this observation was acquired at a rotational phase at which the cloud is behind the star. The inability to detect the cloud is therefore perfectly consistent with the model. 

The other observation of note is the one acquired on RJD 60082. In this case, the RV matches expectations, and the cloud's y-position matches the model, but its x-position is offset to the wrong side of the star. While the error ellipse is close to the expected position of the cloud, it is not quite in 1$\sigma$ agreement. This observation has the worst quality of the dataset, with only three useful baselines and only the DPHI interferometric observable. It is therefore unsurprising that it does not agree as well with the model as the other five datapoints in which the cloud's position could be detected. Notably, the higher-quality observation on 60112 was obtained at a similar rotation phase, and is in much better agreement with the expected cloud position. 

To examine the influence of different modeling assumptions on the astrometric position of the cloud, the bottom right panel of Fig.\ \ref{fig:rhoopha_spectroinf_clouds} shows the variations in the cloud's central positions and the sizes and position angles of the error ellipses for four different cloud models: uniform disks with radii of 0.2, 0.3, and 0.4\,mas, and a Gaussian with a FWHM of 0.2\,mas. These changes have essentially no impact. 

\subsection{\spec{He}{i}{21126}}

The \spec{He}{i}{21126} line shows an absorption line profile, and a weak signal in DPHI at the two epochs corresponding to the largest angular separation between the binary components (RJDs 60053 and 60139). The DPHI signature can be reproduced by introducing a Gaussian line profile for each component. The line profile is defined in the PMOIRED geometrical model in the same way as described above for Br$\gamma$, but with the central wavelength of $2.112583$\,$\mu$m.

The results from the two epochs indicate that the Ab component has a slightly stronger absorption, consistent with it being a He-strong magnetic chemically peculiar star. The RVs of the two components remain similar to those determined from Br$\gamma$, albeit with much larger uncertainties, and therefore are not useful to further constrain the binary orbit.


\section{Discussion}\label{sect:discussion}

\subsection{Comparison to previous results}\label{subsec:result_compare}

\citet{shultz_et_al_2025} determined the orbital elements of $\rho$\,Oph\,A solely from RV measurements. As can be seen from Table \ref{tab:orbital_solution}, the results found here are in agreement in all particulars within the (generally much larger) uncertainties accompanying the original solution. The value of $T_0$ given in the third column of Table \ref{tab:orbital_solution} is almost exactly 75 orbital cycles removed from the more precise period determined by astrometric modeling, and is therefore functionally equivalent. 

Shultz et al.\ additionally derived a distance of $139 \pm 3$~pc, using the GAIA Data Release 3 parallaxes for the nearby stars in the $\rho$\,Oph cluster, as they judged the parallax for $\rho$\,Oph\,A itself to be unreliable due to its high Renormalized Unit Weight Error (RUWE) of 3.4 (where ${\rm RUWE} > 1$ is considered to be indicative of a poor fit due to binarity). The distances determined astrometrically from the orbital fit are in almost perfect agreement with this value. 

\subsection{Spin-orbit alignment}

Shultz et al.\ noted that their orbital inclination $i_{\rm orb} = 69 \pm 11^\circ$ was consistent with the rotational inclination $i_{\rm rot} = 73 \pm 5^\circ$. These values were independently determined: $i_{\rm orb}$ was found from comparison of the mass function to evolutionary models and the locations of the two components on the Kiel diagram, whereas $i_{\rm rot}$ was found from the stellar radius, the rotational period, and the velocity broadening of absorption lines. 

The astrometric value of $i_{\rm orb}$ determined here, $71.31 \pm 0.07^\circ$ (orbfit-lib solution), is much more precise than the previous value, and is functionally identical to $i_{\rm rot}$, suggesting that the orbital and rotational axes of $\rho$\,Oph\,Ab are indeed in perfect alignment. Consistent with this, the astrometric comparison of the expected vs. measured position of the magnetospheric cloud shown in Fig.\ \ref{fig:rhoopha_spectroinf_clouds} provides a very good match if the rotational axis is tilted in the plane of the sky to match $\Omega$, as would be expected for spin-orbit alignment. 

In addition to alignment of the orbital and rotational axes, comparison of the astrometric positions to the RVs of the clouds reveals that $\rho$\,Oph\,Ab also exhibits prograde rotation: both orbit and rotation are counter-clockwise, with maximum and minimum RVs occurring at the south-east and north-west, respectively.

$\rho$\,Oph\,Ab therefore joins the still very limited sample of stars for which the inclination of the rotation axis from the line of sight, the tilt of the rotational axis in the sky, and the direction of rotation have been determined. Similar constraints have sometimes been possible using the starspot tracking method in eclipsing binaries, finding a variety of systems exhibiting aligned, misaligned, prograde, and retrograde rotation \citep[e.g.][]{2015MNRAS.448..429B,2022PASP..134h2001A,2024MNRAS.527.3982L}. Interferometric methods have placed similar constraints on the nearby classical Be star and a wide eccentric binary Achernar \citep[which exhibits prograde rotation but a misaligned rotation axis;][]{2014A&A...569A..10D,2022A&A...667A.111K}, as well as for the Be + stripped star binaries HD\,6819 and HR\,2142, which both exhibit aligned and prograde rotation, as would be expected given that these are post-mass-transfer binaries \citep{2025A&A...694A.208K,2024ApJ...962...70K}. 

\subsection{Formation of the system and origin of the fossil magnetic field}

The almost perfect alignment of the orbital and rotational angular momentum vectors is easily explained in a formation scenario in which both stars were born from the same molecular cloud, and therefore inherited the cloud's angular momentum vector. However, it may challenge the hypothesis that fossil magnetic fields are generated by binary mergers \citep[e.g.][]{2019Natur.574..211S}. 

A merger scenario would need to begin with a hierarchical triple, consisting of the non-magnetic 10 \Msun~primary and two $\sim$4~\Msun~stars, with the smaller stars in a very close orbit. For the smaller stars to merge, their orbit would need to decay, which could happen by one of three mechanisms: 1) stellar evolution, in which one of the stars overflows its Roche lobe; 2) tidal friction; or 3) dynamical interactions. Given the system's age, 1) is ruled out a priori. Tidal friction between B-type stars is inefficient over such timescales, and is unlikely to shrink the orbit sufficiently to produce a merger \citep{1977A&A....57..383Z}. The remaining option, dynamical interactions, would likely be very chaotic. Assuming that the orbit around the primary is not destroyed by the energy released by the merger, the result should still be a much wider and more highly eccentric orbit than is observed \citep{2025A&A...693A..14B}. Furthermore, the perfect alignment of the merger product's angular momentum vector with the orbit is also very unlikely in this scenario. 

The difficulty in reconciling the orbital and rotational properties of the system with a merger suggests that the fossil magnetic field must have some other origin. Here there are two possibilities. The first is that the magnetic flux is simply inherited from the molecular cloud, frozen and amplified during the collapse of the star. High-resolution simulations suggest that this scenario is plausible, but the results are highly sensitive to numerical resolution \citep{2022MNRAS.511..746W}. In the case of binary systems such as $\rho$\,Oph\,A this scenario raises the question of why only one of the stars is magnetic, and might require turbulent processes in the cloud preferentially gathering magnetic flux into one of the cores. Another difficulty concerns the role of convection during the protostellar phase, which should rapidly destroy magnetic flux inherited from the protostellar cloud, unless the rapid formation of a radiative core, or rapid transition to a fully radiative star, can preserve some of the inherited flux \citep{2023A&A...678A.204S}. 

The other non-merger possibility is that the fossil field was generated by a rotational-convective dynamo during the protostellar phase. The principle difficulty with this scenario is that the magnetic activity cycles expected in convective cores should have much shorter timescales \citep[$\sim$ years, as shown by MHD simulations of the convective cores of A-type stars;][]{2024A&A...691A.326H} than the timescale over which radiation freezes out convection, meaning that the radiative zone should end up consisting of thin layers with opposite magnetic polarities that cancel one another out \citep[as described in Sect.\ 4.5 of the review by][]{2017RSOS....460271B}. \cite{2021ApJ...923..104J} have suggested just this sort of magnetic layering as a consequence of dynamo action in the opacity-bump convection zones embedded in the radiative envelopes of early-type stars. Putting this difficulty aside, it is interesting to note that the magnetic secondary is more rapidly rotating than the primary. Assuming that the primary's rotational axis shares the secondary's perpendicular orientation to the orbital plane, its measured velocity broadening of $206 \pm 5$~{\kms} implies a rotational period of about 1 day, indeed slower than the magnetic secondary. Given that the primary is not magnetic, it is unlikely to have experienced significant spindown since arriving on the zero-age main sequence, i.e.\ its present rotational period is probably comparable to its original period. By contrast, the magnetic secondary should be losing angular momentum rapidly through its corotating magnetosphere \citep[e.g.][]{ud2009,petit2013}. Its calculated spindown age of 5 Myr matches the estimated age of the $\rho$\,Oph group, suggesting that it was indeed rotating close to critical velocity when it arrived on the zero-age main sequence. 

If $\rho$\,Oph\,Ab was initially a much more rapid rotator than the primary, and this dichotomy extended back to the system's earliest history, this might explain the magnetic field. More rapid rotation on the pre-main sequence would imply a more vigorous rotational-convective dynamo, and thus a stronger magnetic field, than possessed by the primary; when the stars' envelopes became radiative, the weaker magnetic field of the primary was quickly destroyed by flux decay, while the stronger magnetic field of the secondary was able to freeze itself into the radiative zone. One implication of this scenario is that in young magnetic binary systems, we should generally expect the magnetic star to be the more rapid rotator, while in more evolved systems the magnetic star should rotate more slowly than the non-magnetic companion due to magnetic braking. 

\section{Conclusions}\label{sect:conclusion}

Orbital elements of the magnetic binary $\rho$\,Oph\,A constrained with VLTI/GRAVITY observations confirm the spectroscopically inferred binarity of the system, as well as the gross properties of the orbit (period, semi-major axis, eccentricity, etc.). The orbital inclination $i_{\rm orb} = 71.31 \pm 0.07^\circ$ is identical within uncertainty to the rotational inclination $i_{\rm rot} = 73 \pm 5 ^\circ$ of the magnetic star $\rho$\,Oph\,Ab. 

We obtained a highly significant detection of the dominant magnetospheric cloud orbiting $\rho$\,Oph\,Ab from the spectro-interferometric data, with the strongest signal appearing in the differential phases. The detection and the astrometric measurement of its changing position around the magnetic secondary furthermore demonstrates that 1) the magnetic star's rotational axis is tilted in the plane of the sky by the same angle as the longitude of the ascending node, i.e. the rotational axis is perpendicular to the orbital plane, and 2) that the magnetic star exhibits prograde rotation. 

This is the first time that a hot star magnetosphere has been detected astrometrically, and therefore the first case in which the direction of a magnetic early-type star's rotation has been determined. Astrometric detection of the magnetosphere additionally provides further confirmation of the gap between the inner edge of the magnetically confined plasma of a circumstellar magnetosphere (with an inner boundary at the Kepler corotation radius) and the surface of the star, as expected from the Rigidly Rotating Magnetosphere model. 

\section*{Data availability}

The reduced and calibrated GRAVITY data are available at \url{https://oidb.jmmc.fr/collection.html?id=512015b1-fb46-4cde-a015-b1fb46ccdec6}.

\begin{acknowledgements}
This work is based on observations collected at the European Southern Observatory under ESO programme 111.24M8.001. MES acknowledges financial support from the ESO SSDF. This research has made use of the Jean-Marie Mariotti Center \texttt{Aspro} service.
\footnote{Available at http://www.jmmc.fr/aspro} This research has made use of the Jean-Marie Mariotti Center \texttt{OIFits Explorer} service. 
\footnote{Available at http://www.jmmc.fr/oifitsexplorer} This research has made use of the Jean-Marie Mariotti Center \texttt{SearchCal} service, which involves the JSDC and JMDC catalogues.
\footnote{Available at https://www.jmmc.fr/searchcal}
\end{acknowledgements}

\bibliographystyle{aa} 
\bibliography{bibliography} 

\onecolumn

\begin{appendix}

\section{Detailed results from the fitting of GRAVITY data}\label{appendix}

\begin{figure*}[h!]
\begin{center}
\includegraphics[width=16.5cm]{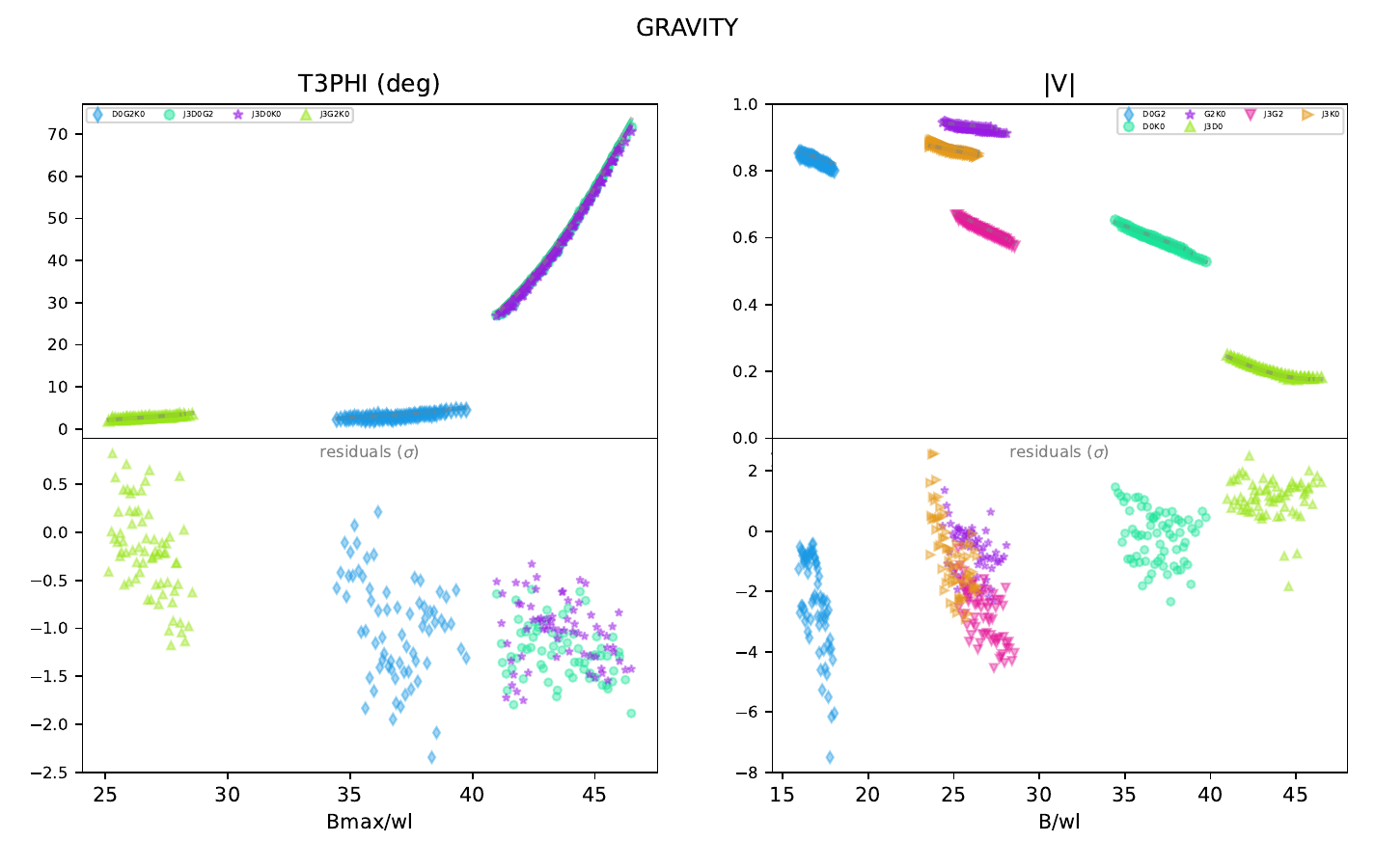}%
\end{center}
\caption[xx]{\label{fig:binary_fit_2023Sep17} {\em Top}: Binary fit to the rebinned GRAVITY data taken on RJD 60205. The data consist of four sets of closure phases (T3PHI in the unit of degrees) for each telescope triangle and six sets of visibilities (|V|) for each telescope pair (colored points). The data are overplotted with the binary model fit (dashed lines). {\em Bottom}: The residuals in units of $\sigma$. The x axes correspond to the spatial frequency in the unit of M$\lambda$. The binary fit indicates a relative offset $\rho\sim2.3$\,mas, PA$\sim$223$^{\circ}$, and $f=f_{\rm Ab}/f_{\rm Aa}\sim0.7$. The reduced $\chi^2$ of the fit is $\sim2.7$. 
}
\end{figure*}

\begin{figure*}[]
\begin{center}
\includegraphics[width=17cm]{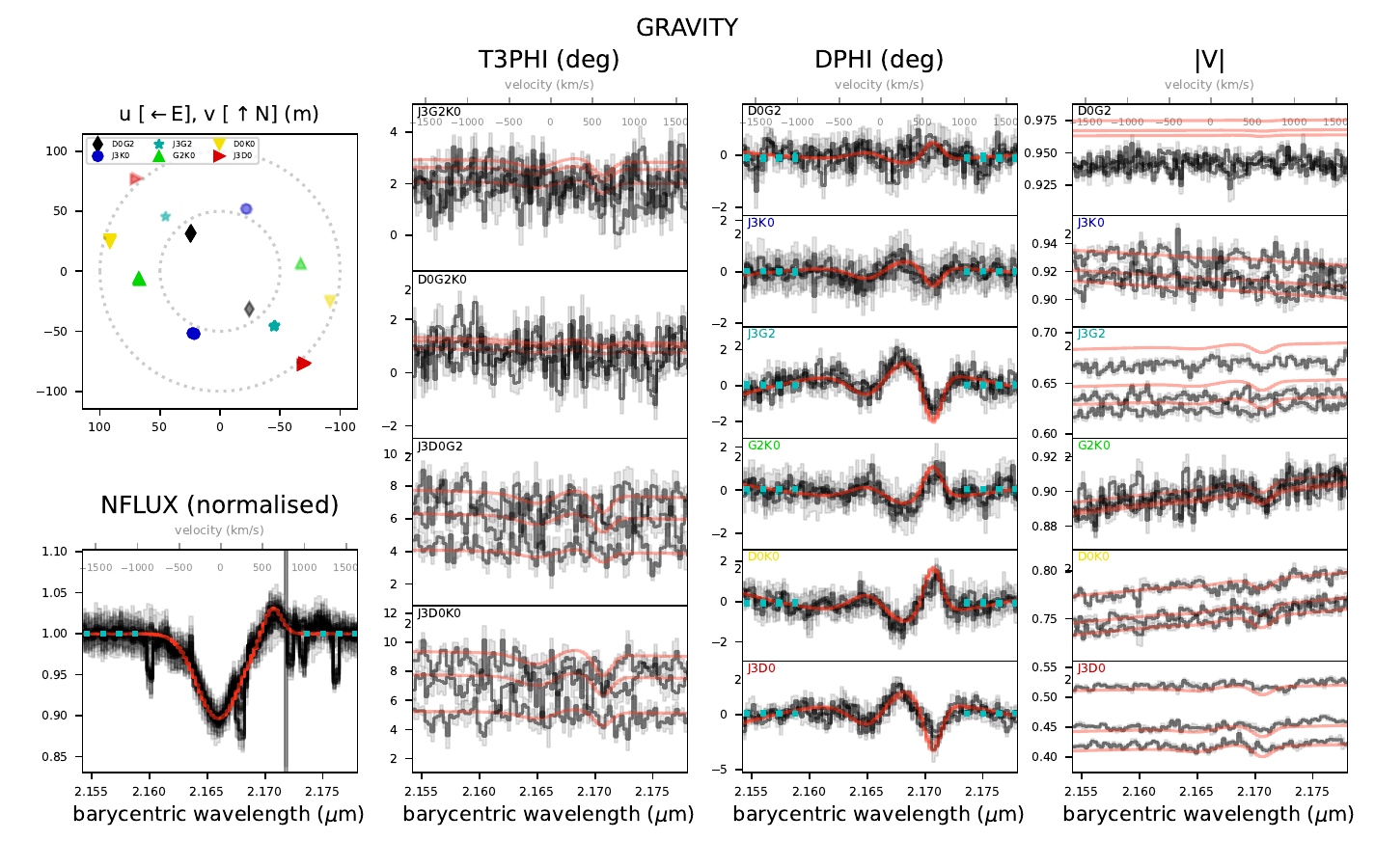}%
\end{center}
\caption[xx]{\label{fig:linefit_2023Apr18} Same as Fig.~\ref{fig:linefit} but for RJD 60054.}
\end{figure*}

\begin{figure*}[]
\begin{center}
\includegraphics[width=17cm]{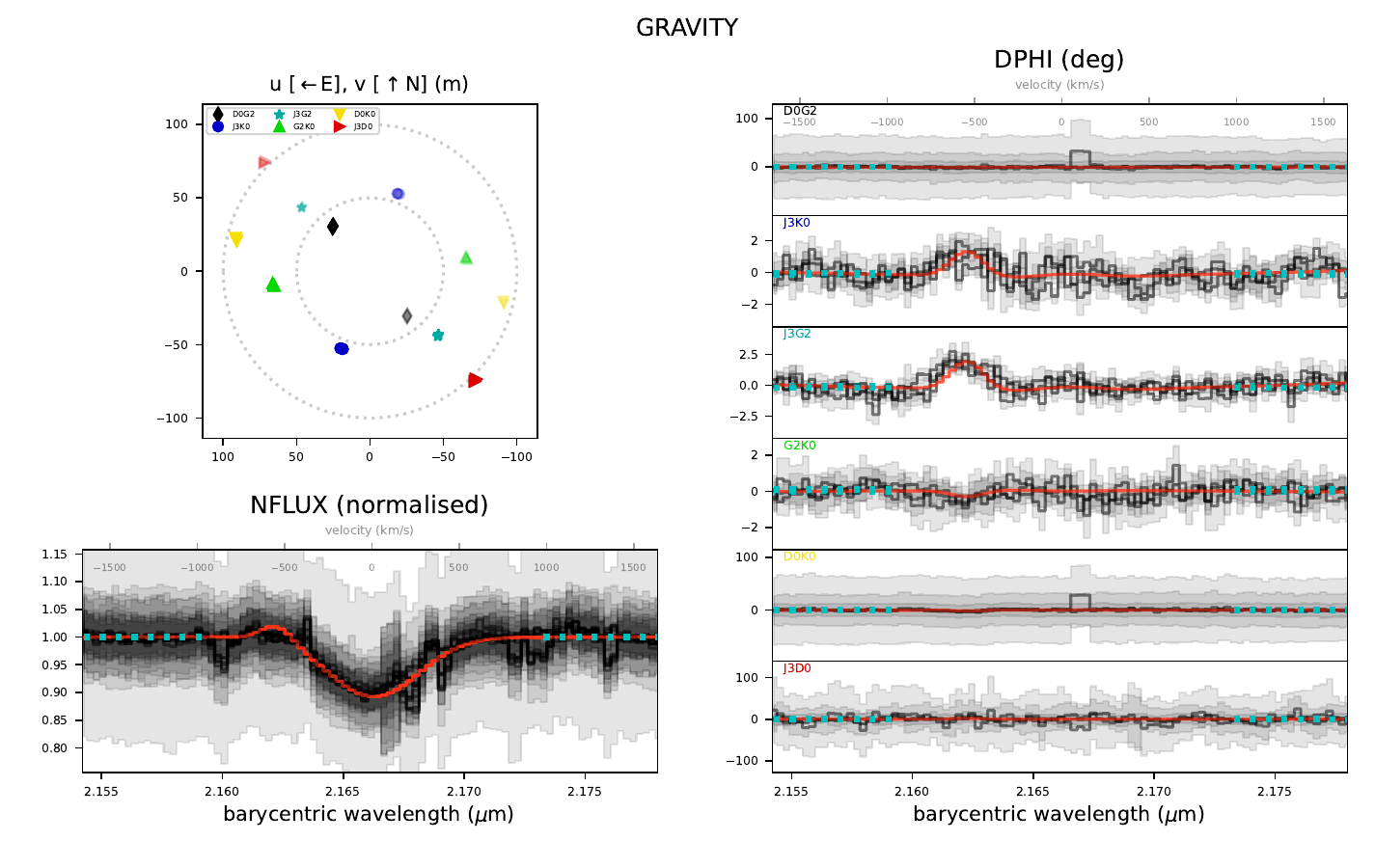}%
\end{center}
\caption[xx]{\label{fig:linefit_2023May16} Same as Fig.~\ref{fig:linefit} but for RJD 60082 and not showing the discarded T3PHI and |V| observables.}
\end{figure*}

\begin{figure*}[]
\begin{center}
\includegraphics[width=17cm]{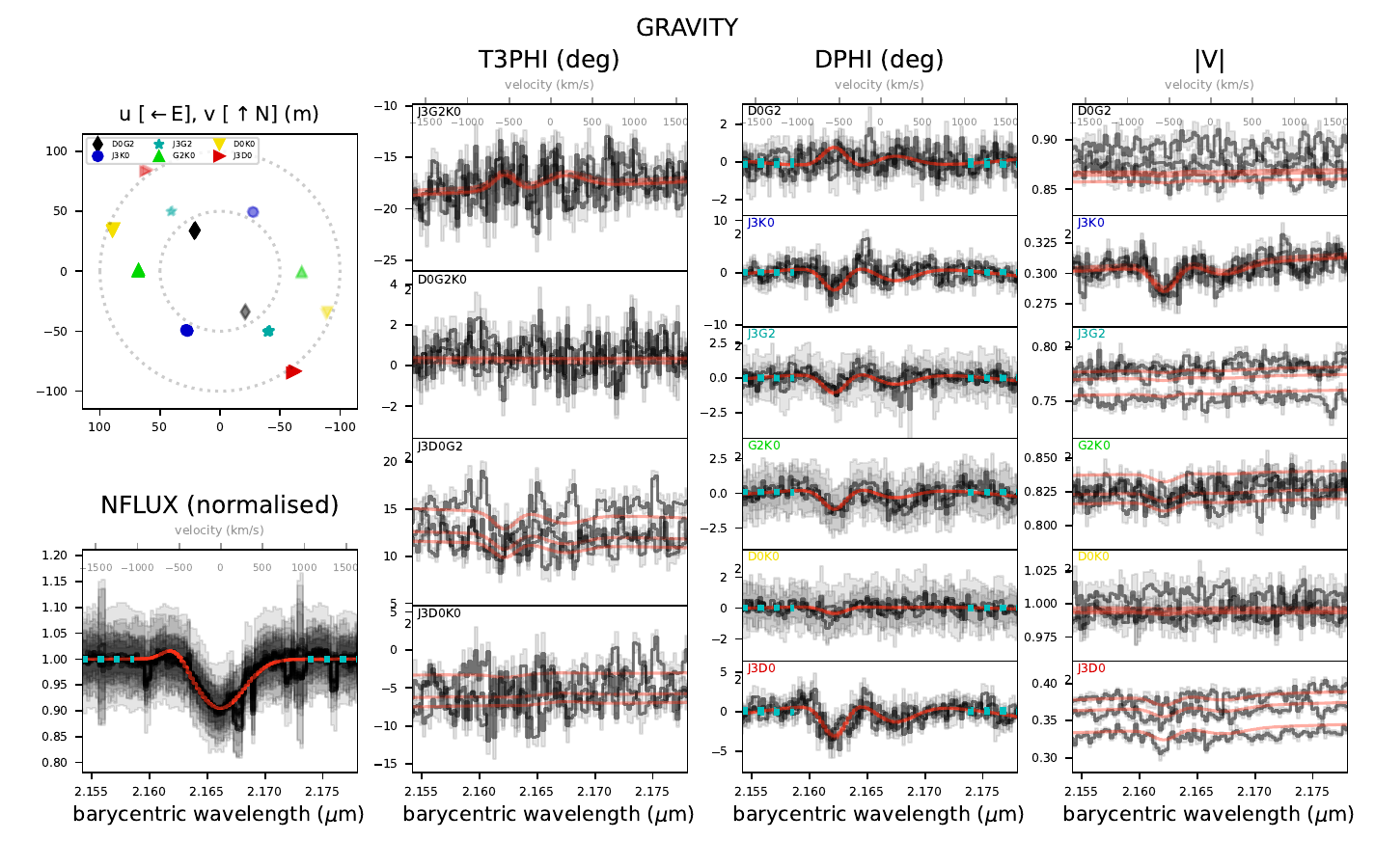}%
\end{center}
\caption[xx]{\label{fig:linefit_2023Jun15} Same as Fig.~\ref{fig:linefit} but for RJD 60112.}
\end{figure*}

\begin{figure*}[]
\begin{center}
\includegraphics[width=17cm]{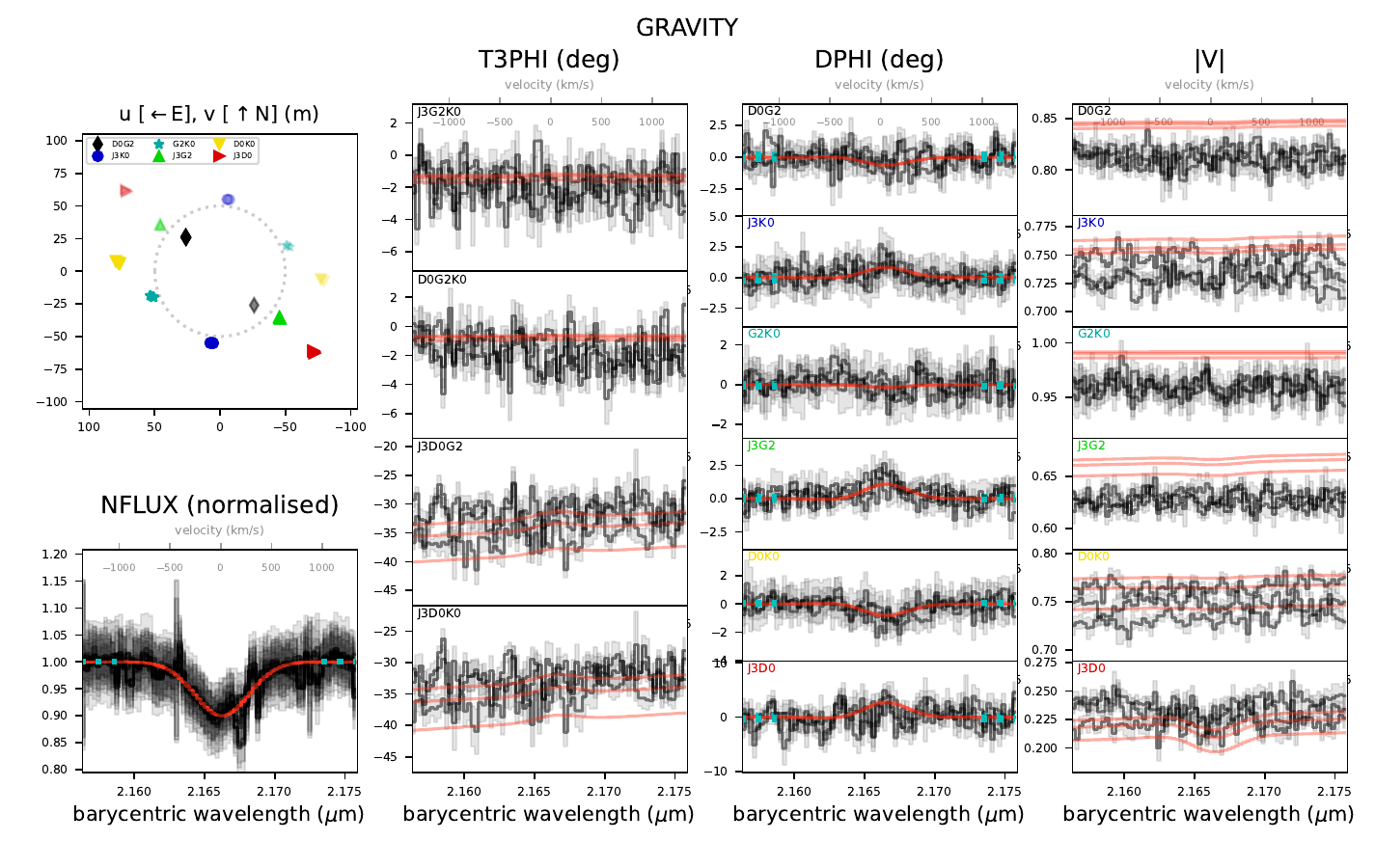}%
\end{center}
\caption[xx]{\label{fig:linefit_2023Aug13} Same as Fig.~\ref{fig:linefit} but for RJD 60171.}
\end{figure*}

\begin{figure*}[]
\begin{center}
\includegraphics[width=17cm]{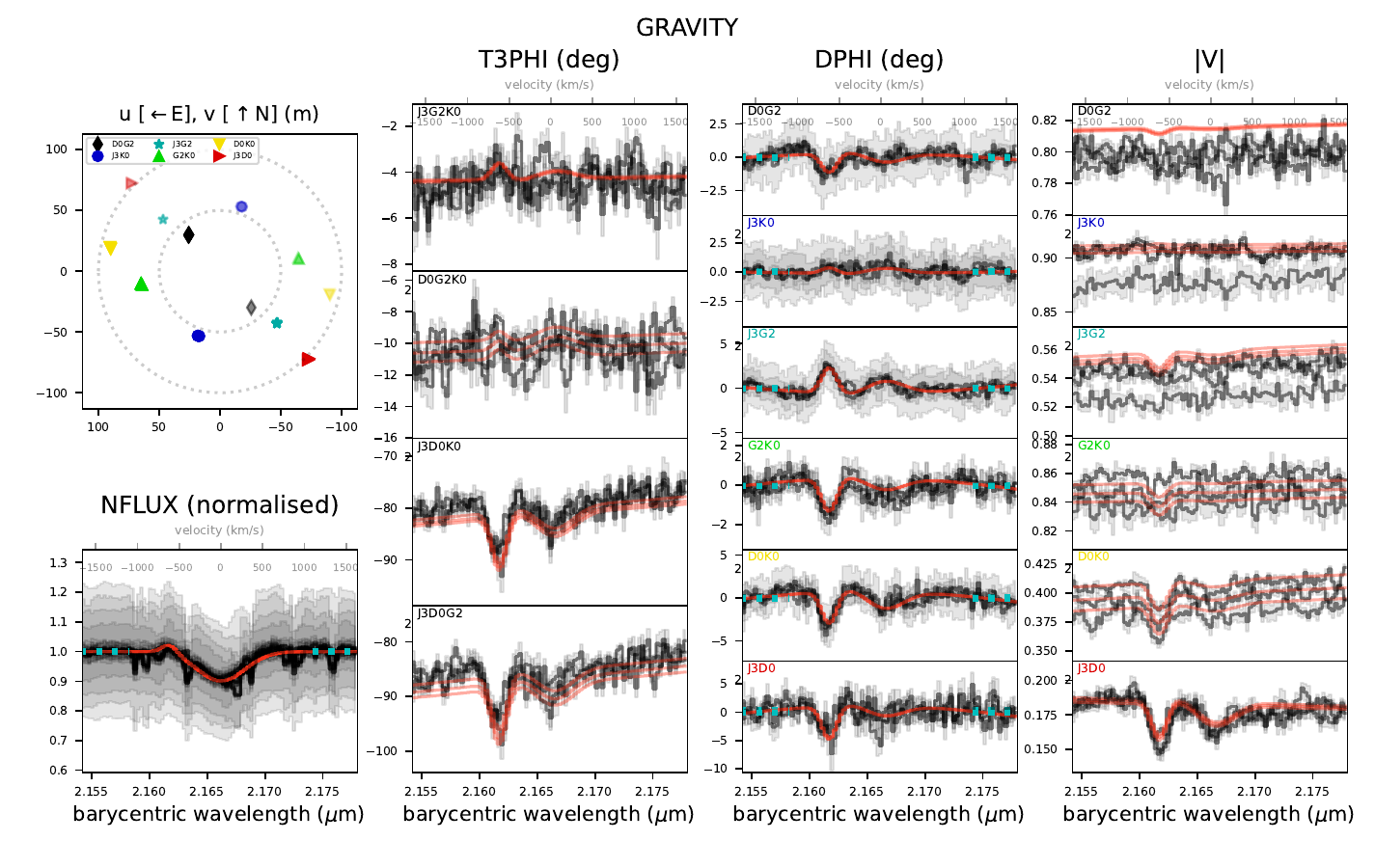}%
\end{center}
\caption[xx]{\label{fig:linefit_2023Aug14} Same as Fig.~\ref{fig:linefit} but for RJD 60172.}
\end{figure*}

\begin{figure*}[]
\begin{center}
\includegraphics[width=17cm]{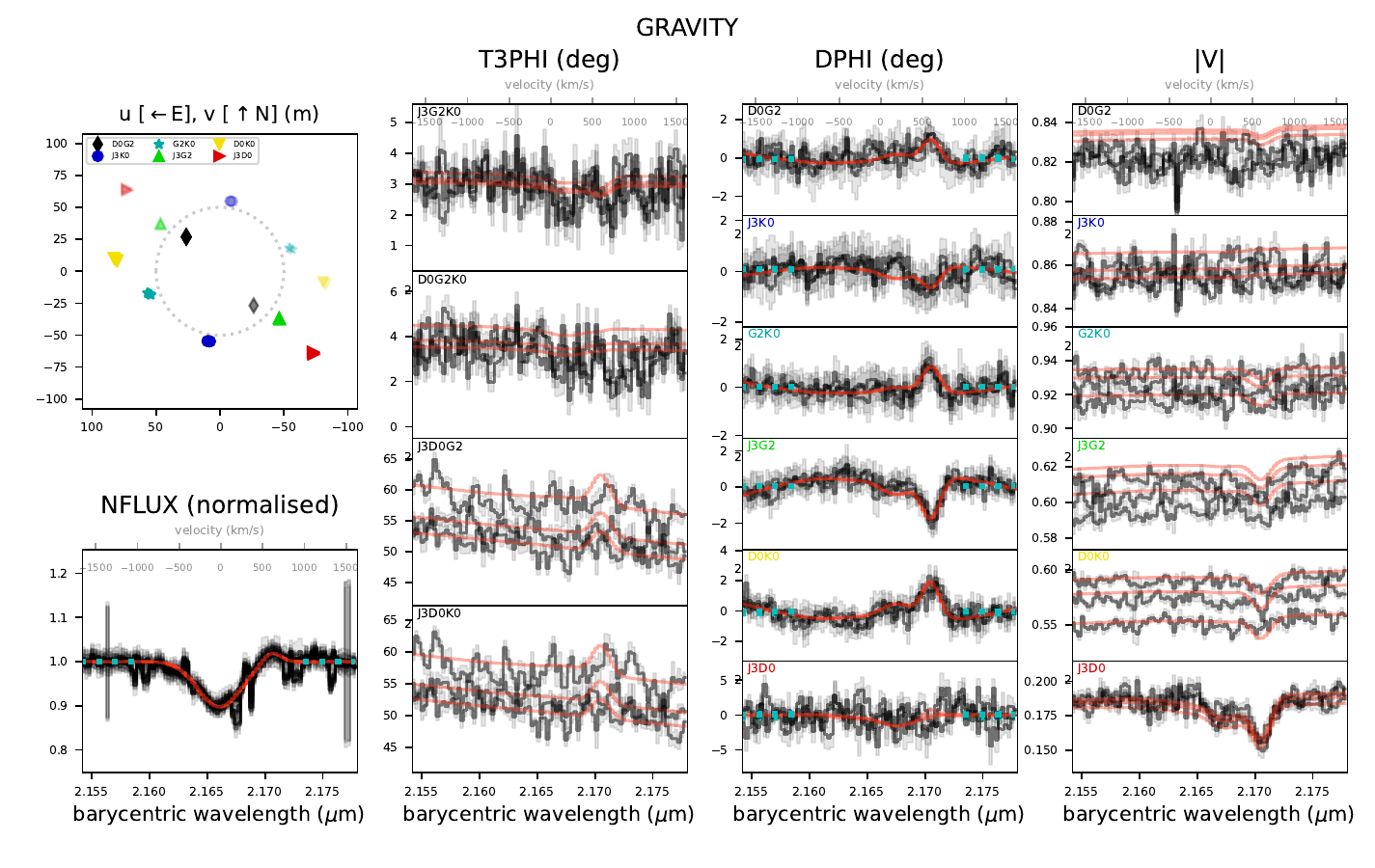}%
\end{center}
\caption[xx]{\label{fig:linefit_2023Sep17} Same as Fig.~\ref{fig:linefit} but for RJD 60205.}
\end{figure*}

The relative astrometry and flux ratios for all the GRAVITY epochs are detailed in Table~\ref{tab:astrometry}, while the RVs for the stellar components and the RVs and the astrometric positions of the magnetospheric cloud derived from the GRAVITY spectro-interferometry are given in Table~\ref{tab:gravity_rvs_cloud_astrometry}. An example of the binary model fit to GRAVITY data is shown in Fig.~\ref{fig:binary_fit_2023Sep17}, while the fits to the Br$\gamma$ spectro-interferometry are shown in Figs.~\ref{fig:linefit_2023Apr18} to \ref{fig:linefit_2023Sep17} (except for RJD 60139 which is shown in Fig.~\ref{fig:linefit}).

\begin{table*}[h!]
\caption{Relative astrometric positions and flux ratios of the $\rho$ Oph A binary determined from GRAVITY data}         
\label{tab:astrometry}      
\centering          
\begin{tabular}{l c c c c c c c c c c} 
\hline\hline       
Date & RJD & Baselines & $\rho$ & PA & $\Delta$RA & $\Delta$DEC & $\sigma$-$a$ & $\sigma$-$b$ & $\sigma$-PA & $f=f_{\rm Ab}/f_{\rm Aa}$ \\ 
 &  & & [mas]  & [$^\circ$] & [mas] & [mas] & [mas] & [mas] & [$^\circ$] &  \\
\hline
2023\,Apr\,18 & 60053.836 & D0-G2-J3-K0 & 8.449 & 301.670 & -7.191 & 4.436 & 0.0077 & 0.0026 & 173.6 & 0.709$\pm$0.005\\
2023\,May\,16 & 60081.777 & D0-G2-J3-K0 & 2.317 & 20.407 & 0.808 & 2.172 & 0.0494 & 0.0078 & 139.4 & 0.576$\pm$0.030\\
2023\,Jun\,15 & 60111.637 & D0-G2-J3-K0 & 3.315 & 157.030 & 1.294 & -3.052 & 0.0047 & 0.0039 & 113.1 & 0.702$\pm$0.002\\
2023\,Jul\,12 & 60138.628 & D0-G2-J3-K0 & 8.169 & 298.888 & -7.153 & 3.947 & 0.0029 & 0.0018 & 140.3 & 0.705$\pm$0.001\\
2023\,Aug\,13 & 60170.598 & D0-G2-J3-K0 & 2.305 & 29.272 & 1.127 & 2.010 & 0.0242 & 0.0082 & 139.8 & 0.700$\pm$0.011\\
2023\,Aug\,14 & 60171.534 & D0-G2-J3-K0 & 2.266 & 42.504 & 1.531 & 1.671 & 0.0054 & 0.0023 & 169.7 & 0.703$\pm$0.002\\
2023\,Sep\,17 & 60205.487 & D0-G2-J3-K0 & 2.252 & 222.747 & -1.529 & -1.654 & 0.0050 & 0.0023 & 151.2 & 0.699$\pm$0.003\\
\hline
\end{tabular}
\tablefoot{$\rho$ is the angular separation between the components, PA is the position angle (from north to east), $\Delta$RA and $\Delta$DEC are the Ab coordinates relative to Aa in the East and North direction, respectively, $\sigma$-$a$ and $\sigma$-$b$ are the major and minor axes of the error ellipse, respectively, $\sigma$-PA is the position angle of the error ellipse (from north to east), and $f=f_{\rm Ab}/f_{\rm Aa}$ is the flux ratio.\\
}
\end{table*}

\begin{table*}[h!]
\caption{RVs of the stellar components and RVs and astrometric positions of the magnetospheric cloud determined from GRAVITY data}         
\label{tab:gravity_rvs_cloud_astrometry}      
\centering          
\begin{tabular}{l c c c c c c c c c c} 
\hline\hline       
RJD & RV$_{\rm Aa}$ & RV$_{\rm Ab}$ & RV$_{\rm cloud}$ & $\Delta$RA & $\Delta$DEC & $\sigma$-$a$ & $\sigma$-$b$ & $\sigma$-PA \\ 
 & [$\kms$] & [$\kms$] & [$\kms$] & [mas] & [mas] & [mas] & [mas] & [$^\circ$]  \\
\hline
60053.836 & $-51.9\pm1.7$ & $35.9\pm4.3$   & $621.0\pm2.3$   & -0.212 & 0.091 & 0.149 & 0.039 & 176.3 \\
60081.777 & $23.3\pm3.3$  & $11.2\pm5.3$   & $-522.6\pm13.6$ & -0.107 & -0.161 & 0.185 & 0.089 & 170.8 \\ 
60111.637 & $30.0\pm2.3$  & $-42.1\pm12.7$ & $-544.7\pm14.4$ & 0.253  & -0.196 & 0.118 & 0.062 & 138.7 \\ 
60138.628 & $-47.2\pm1.3$ & $71.0\pm2.8$   & $-569.0\pm2.8$  & 0.381  & -0.344 & 0.091 & 0.023 & 135.7 \\
60170.598\tablefootmark{a} & $28.0\pm4.7$  & $8.9\pm5.8$   & $-14.6\pm230$   & - & - & - & - & - \\ 
60171.534 & $30.3\pm1.8$  & $0.3\pm5.2$    & $-609.1\pm2.5$  & 0.392 & -0.157 & 0.287 & 0.062 & 172.0 \\
60205.487 & $8.0\pm2.5$   & $-32.9\pm5.1$  & $600.6\pm3.2$   & -0.456 & 0.105 & 0.129 & 0.046 & 143.7 \\
\hline
\end{tabular}
\tablefoot{RVs for Aa, Ab, and the magnetospheric cloud are listed with errors from bootstrapping, to which we quadratically added the estimated GRAVITY wavelength calibration uncertainty of $0.05$\% ($15\,\kms$). $\Delta$RA and $\Delta$DEC are the astrometric offsets of the magnetospheric cloud relative to the magnetic Ab component in the East and North direction, respectively, $\sigma$-$a$ and $\sigma$-$b$ are the major and minor axes of the error ellipse, respectively, and $\sigma$-PA is the position angle of the error ellipse (from north to east).\\
\tablefoottext{a}{The magnetospheric cloud was eclipsed by the Ab component at this epoch, and its astrometric position and RV thus could not be fitted. For the RV, we adopted the systemic velocity with the stellar $v\sin{i}$ as the uncertainty (see text).}
}
\end{table*}

\end{appendix}

\end{document}